# RAPID CYCLING SYNCHROTRON OPTION FOR PROJECT X[*]

WEIREN CHOU[†]

*Fermi National Accelerator Laboratory*
*P.O. Box 500, Batavia, Illinois 60510, USA*

This paper presents an 8 GeV Rapid Cycling Synchrotron (RCS) option for Project X. It has several advantages over an 8 GeV SC linac. In particular, the cost could be lower. With a 2 GeV 10 mA pulsed linac as injector, the RCS would be able to deliver 4 MW beam power for a muon collider. If, instead, a 2 GeV 1 mA CW linac is used, the RCS would still be able to meet the Project X requirements but it would be difficult for it to serve a muon collider due to the very long injection time.

## 1. Introduction

Work on an RCS as an option for a proton driver for a muon collider began around 1998. Several laboratories including Fermilab, BNL, KEK and RAL carried out studies and published design reports. At Fermilab, two proton driver studies, with acronyms PD1 and PD2, were initiated in the period from 1998 to 2002. PD1 was a 16 GeV RCS and PD2 an 8 GeV RCS. Both studies were documented [1,2]. In 2004, Fermilab decided to abandon the synchrotron approach and opted for an 8 GeV SC linac, also a proton driver. The principal reason for this choice was its synergy with the ILC, the ultimate goal of Fermilab at that time. Unfortunately, the proton driver proposal was rejected by DOE because of the "fast track" strategy for the ILC. In 2007, however, when the ILC cost estimate was published and it was seen to be about twice the previous TESLA estimate, DOE lost enthusiasm for this project. Fermilab decided to go back to the proton driver, which was renamed Project X. Furthermore, DOE allowed decoupling between Project X and the ILC. Thus, after a complete circle the RCS option is back on the table. In the meantime, Japan has built a 1 MW 3 GeV RCS as part of the J-PARC project.

There are two fundamental requirements of a proton driver: high beam power and short bunch length. Both are achievable by either an RCS or a linac.

---

[*] This work is supported by Fermi Research Alliance, LLC under Contract No. DE-AC02-07CH11359 with the United States Department of Energy.
[†] E-mail: chou@fnal.gov

The main advantage of an RCS is the lower cost. A preliminary comparison shows the cost of an 8 GeV RCS could be 40% lower than an 8 GeV SC linac ($0.93B *vs.* $1.5B). Moreover, since the injection energy of an RCS is low, the stripping foil would be easier, stripping efficiency higher and more beam loss at injection could be tolerated. The transport of H¯ ions in the beam line would be easier because stronger bends could be used and there would be no need for a cryogenic beam pipe. Existing enclosures could be reused. On the other hand, however, a linac is simpler to design, build and operate than a synchrotron and has higher reliability. The hardware of an RCS is more challenging (large aperture magnets, rapid cycling power supplies, high power tunable RF system, field tracking during cycle, eddy current in beam pipe, etc.).

**2. RCS for Project X**

Compared to previous RCS studies, there are several new conditions that would make a better RCS possible for Project X.

- Elimination of the Main Injector (MI) injection front porch: This is piggy-backing on the NOvA project, which uses the Recycler as an accumulator.
- Free housing: The antiproton source enclosure would be available after 2011 and could house an 8 GeV RCS (see below).
- Higher injection energy: A 2 GeV SC linac (*vs.* 600 MeV in PD2) as an injector of the RCS would reduce the RCS aperture requirement.

Fig. 1 shows an 8 GeV RCS located in the present antiproton source enclosure. It has the same size and shape as the Debuncher [3]. The triangular lattice shown in Fig. 2 is transition-free thanks to its high $\gamma_t$ (18.6). This is a simple doublet lattice and uses only one type of bending magnets and one type of quadrupoles. The missing magnet scheme in mid-cell provides zero-dispersion straight sections without dispersion suppressors [4].

**3. RCS for the Muon Collider**

To use an RCS as a proton driver for a muon collider, the main requirements are: 4 MW beam power and 3 ns bunch length. We consider two scenarios.

**3.1.** *Pulsed Linac*

Table 1 lists the parameters when a 10 mA pulsed linac is used as the RCS injector. In order to reduce the bunch number from 90 to 5, one may add a compressor ring, which would coalesce 18 bunches into one and form 5 super-bunches, each of $2 \times 10^{13}$ protons (20 TP). These super-bunches would be phase-rotated to 3 ns in length, a compression ratio of 6. The bunch spacing can be controlled by using RF barriers.

### 3.2. *CW Linac*

Table 1 also lists the parameters when a 1 mA CW linac is used as the RCS injector. It is seen that the injection time becomes very long (16 ms), which would lead to two major problems. One is the excessive heating on the stripping foil due to the large number of hits per particle. Another is 30 Hz operation of the RCS would be questionable because the injection takes half cycle time.

It should be pointed out that a CW linac serving the RCS would be fine for Project X, since the RCS needs only to deliver 0.5 MW, not 4 MW.

Table 1. Main parameters of a 4 MW RCS with a pulsed or CW linac.

|  | **Pulsed Linac** | **CW Linac** |
|---|---|---|
| **Linac** | | |
| Rep rate (Hz) | 30 | CW |
| Kinetic energy (GeV) | 2 | 2 |
| Peak current (mA) | 10 | 1 |
| Pulse length (µs) | 1600 | 16000 |
| H⁻ per pulse | $1 \times 10^{14}$ | $1 \times 10^{14}$ |
| Beam power (MW) | 1 | 1 |
| **RCS** | | |
| Rep rate (Hz) | 30 | 30 |
| Circumference (m) | 505.3 | 505.3 |
| Extraction energy (GeV) | 8 | 8 |
| Protons per bunch | $1.1 \times 10^{12}$ | $1.1 \times 10^{12}$ |
| Number of bunches | 90 | 90 |
| Protons per cycle | $1 \times 10^{14}$ | $1 \times 10^{14}$ |
| Norm. transverse emit. (mm-mrad) | $40\pi$ | $40\pi$ |
| Longitudinal emit. (eV-s) | 0.2 | 0.2 |
| Injection time (µs) | 1600 | 16000 |
| Injection turns | 950 | 9500 |
| Maximum Laslett tune shift | 0.15 | 0.15 |
| Extraction bunch length (rms, ns) | 1 | 1 |
| RF frequency (MHz) | 53 | 53 |
| Beam power (MW) | 4 | 4 |

**References**

1. "The Proton Driver Design Study," *Fermilab-TM-2136* (December 2000).
2. "Proton Driver Study II – Part 1," *Fermilab-TM-2168* (May 2002).
3. W. Chou, "A Hybrid Design of Project X," *PAC2009 Proceedings*, May 4-8, 2009, Vancouver, Canada.
4. W. Chou, "A Simple Transition-free Lattice of an 8 GeV Proton Synchrotron," *PAC2009 Proceedings*, May 4-8, 2009, Vancouver, Canada.

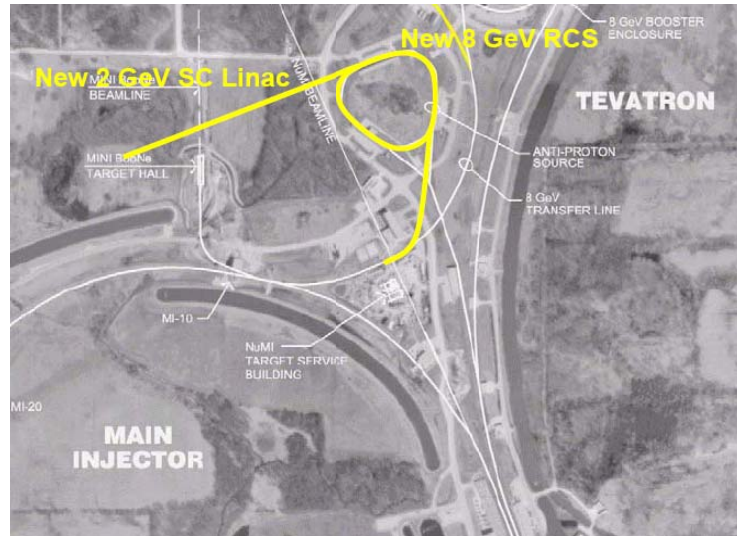

Figure 1. Layout of a 2 GeV SC linac and an 8 GeV RCS.

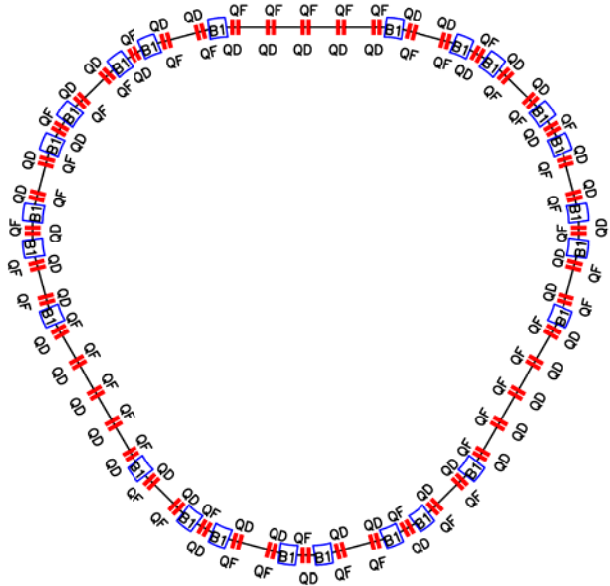

Figure 2. 2D plot of a triangular transition-free lattice.